# Data as Infrastructure for Smart Cities: Linking Data Platforms to Business Strategies

Larissa Romualldo-Suzuki, Anthony Finkelstein

*a Department of Computer Science, University College London, WC1E 6BT, United Kingdom*

**Abstract**

The systems that operate the infrastructure of cities have evolved in a fragmented fashion across several generations of technology, causing city utilities and services to operate sub-optimally and limiting the creation of new value-added services and restrict opportunities for cost-saving. The integration of cross-domain city data offers a new wave of opportunities to mitigate some of these impacts and enables city systems to draw effectively on interoperable data that will be used to deliver smarter cities. Despite the considerable potential of city data, current smart cities initiatives have mainly addressed the problem of data management from a technology perspective, have treated it as a single and disjoint ICT development project, and have disregarded stakeholders and data needs. As a consequence, such initiatives are susceptible to failure from inadequate stakeholder input, requirements neglecting, and information fragmentation and overload. They are also likely to be limited in terms of both scalability and future proofing against technological, commercial and legislative change. This paper proposes a systematic business-model-driven framework, named SMARTify, to guide the design of large and highly interconnected data infrastructures which are provided and supported by multiple stakeholders. The framework is used to model, elicit and reason about the requirements of the service, technology, organization, value, and governance aspects of smart cities. The requirements serve as an input to a closed-loop supply chain model, which is designed and managed to explicitly consider the activities and processes that enables the stakeholders of smart cities to efficiently leverage their collective knowledge at R&D, Procurement, Roll-Out and Market stages. We demonstrate how our approach can be used to design data infrastructures by examining a series of exemplary scenarios and by demonstrating how our approach handles the holistic design of a data infrastructure and informs the decision making process.

## 1. Introduction

In the past four years, organizations, citizens and research institutes have become adept at holding Governments to account for the environmental, social and economic consequences of population growth (Caragliu et al., 2009; Dirks et al., 2010; Hollands, 2010). Since then, cities have been ranked on the basis of the "smartness" of their smart cities strategy, data and technological infrastructures (e.g. Giffinger, 2007, Global Open Data Index[1]) and, despite sometimes divergent and questionable methodologies, these rankings attract considerable publicity and marketing of technology products. As a result, *smart cities* have emerged as an inescapable priority for policy makers in every city in the world.

This label has been applied to a range of technologies powered by data that can assist cities to plan for population growth, and introduce a more sustainable, efficient, and liveable model in urban development. Governments tie the success of their smart cities initiatives directly to the achievement of positive effects for humankind, mobility, urban services and the natural environment.

Many cities and local governments alongside technology organizations and research institutes have already done work to link smart cities solutions to policy goals and initiatives. The availability of large-scale cross domain city data has the potential to drive economic growth, improve government transparency, and create innovative new value-added city services (Kleinman, 2016). Implemented data-driven solutions for smart cities include smart electricity grids (Klein and Kaefer, 2008), public spaces monitoring (Filipponi et al., 2010), and smart buildings (Al-Hader et al., 2009), congestion charging zone[2,] air quality monitoring[3], intelligent road crossings[4]. The majority of these solutions are powered by sensory and open city data provided by many sources within the cities.

---

[1] http://index.okfn.org/

[2] https://tfl.gov.uk/modes/driving/congestion-charge
[3] http://www.londonair.org.uk/LondonAir/Default.aspx
[4] https://tfl.gov.uk/info-for/media/press-releases/2014/march/tfl-to-launch-worldleading-trials-of-intelligent-pedestrian-technology-to-make-crossing-the-road-easier-and-safer

The most prevalent effort powering smart cities solutions are government open data provided in data catalogues websites. Ultimately, the governments envision that their data catalogues will offer city data capable of increasing interoperability between services, security, and public engagement. The UK and the US open data initiatives are good examples of government efforts to contribute valuable information to the constituents.

Yet these efforts have not been nearly as productive as they could be – for at least five reasons. First, governments and policy makers pressure cities to think of smart cities and city data offering in generic ways instead of in the way most appropriate to each city's strategy. Second, they pit city needs against society needs for city data, when clearly the two are interdependent. Third, they focus on built data catalogues rather than platforms that can steer strong value networks of data and services providers. Fourth, they sustain a fragmented data supply chain, hindering city data to be exploited to its full effect. Fifty, due to a lack of a strategy they have been overtaken by technical, logistical and organizational complexity, and their data catalogue initiatives have become fragmented, expensive and difficult to design, maintain and evolve. One of the potential causes for it is that both the technology and (especially) the market are immature. As a result, many cities take the existing infrastructures as their starting point without making much consideration whether the adopted strategy is the best one to follow.

The fact is, the prevailing approaches to city data provision are so fragmented and so disconnected from a strategic and outcomes-oriented approach as to obstruct many of the greatest opportunities for city data to benefit societies and economies. If, instead, cities were to analyse their vision for smart cities using a framework that guide their core strategies, they would discover that infrastructure for city data provision can be much more than an expensive and complex high-tech project – it can be a source of transformational business models, innovation and competitive advantage.

In this paper, we introduce a framework that cities can use to think strategically about how systems, businesses and individuals can draw effectively on interoperable cross-domain city data. When looked at strategically, government data catalogues can be incrementally transformed into data infrastructures, which we define as "*the basic physical, digital, organisational, value and governance structures and processes needed for the management of all data that underpins the decision making processes in smart cities*".

Our framework proposes a new way to look at the relationship between business models and city data that does not treat the development of data catalogues as a disjoint ICT project which are susceptible to failure from inadequate stakeholder input, requirements neglecting, and information fragmentation and overload. Our findings suggest that in contrast to data catalogues, data infrastructures associated with with visualisation and analytical capability become sources of tremendous social change, as cities and stakeholders collaboratively applies their resources, expertise and insights to deliver data that can offer unpreceded opportunities to solve social and environmental challenges.

This framework has been adopted to design urban platforms adopted by 27 European Cities as part of a European Commission project. This work has created the entire definition of goals, the development of use cases, and specifications of services, technology, organisation (stakeholders), value (finance and smart cities impact) and governance domain) for European cities. Ultimately, the adoption of our framework will lead to reduced R&D, procurements times, increased confidence in platform designs to accelerate Roll Out time, design of cost effective and innovative solutions designed with levels of collaboration to reduce data infrastructures time to Market.

## 2. The Emergence of Data Platforms

The systems that operate the infrastructure of cities have evolved in a fragmented fashion across several generations of technology, causing city utilities and services to operate sub-optimally and limiting the creation of new value-added services. These challenges and the scale of city-wide technology adoption have forced cities to rethink their strategies and drive the design of innovative and cost-efficient digital technologies that will provide citizens with a high quality of life while meeting their ambitious sustainability agenda.

With the growing importance of service innovation in smart cities, city data becomes a more important element in the innovation strategy of cities, which means that more capabilities and resources have to be made available. To date, many smart cities solutions have been enabled by city data held by both private and public organizations as well as crowdsourced data. Most of the existing solutions, including government data catalogues, have moved all the way into large corporations such as IBM, Cisco Systems, Accenture, Arup, Siemens, Socrata and Datapress. As of 2012, approximately 143 "self-designated" smart city projects, primarily led by technology organizations, were under development or under completion in the USA, Europe and Asia (Lee and Kwak, 2012). Examples of these are the city of Barcelona and Cisco are embedding sensors on bins to testing whether the routes of refuse collection vans can be optimised by only sending them to full bins. The city estimates that the system could save 10 per cent on waste disposal[5]. Alliander is implementing flexible street lighting during the night in Amsterdam and Glasgow. More recently, governments around the world have provided thousands of government open datasets to offer additional data to address complex urban problems. The intensified provision of Government open data through online platforms has not been entirely voluntary. Many cities, especially in the UK and the US awoke to it only after being surprised by citizens and NGOs demands for increased transparency and accountability (Janssen, 2011; Horsley, 2006; Harrison, 2012).

The UK Government has been strongly committed to become the most open and transparent government in the world. The London Datastore was one of the first platforms to make public data open and accessible, and its role was made clear in the Mayor of London's election manifesto "*Boost London Data store to make the Mayoralty even more transparent and give Londoners access to more information on how I am meeting my pledges*" and "*Encourage more partners to use the data on the Data store to create smartphone apps*". Since its launch it has published over 600 datasets with open data certificates to assure quality, and has led to the creation of more than 200 apps, such as the Citymapper travel app has recently closed a $40m series B investment and has now been exported to some of the biggest cities in the world, although the ROI and long term business model is unclear[6]. The London Datastore has been internationally recognized and its success has earned the Greater London Authority the Open Data Publisher Award offered by the Open Data Institute (ODI).

Shortly after the first releases of Government Open Data (GOD), cities realised that the re-use of their data by private and public bodies became an instrument to foster innovation, and to improve and create new urban services. Open data have been used to plan emergency responses (Jung, 2009), to understand how people move and commute within cities (Gonzalez, 2008), businesses can improve their operation and target audiences, citizens can plan their journeys by different modes of transport (McNamara, 2008), and innovators and start-ups can access data that will help them to create valuable new businesses. Transport for London (TfL), the leading open data provider in the Transport sector in London, provides over 30 open data feeds to over 5,000 application developers design travel applications, tools and services. Their website receives 600,000 unique views on an average day, and 10 million data feed hits a week. Although TfL has no data strategy except to release it, the organisation estimates that in 2013 alone their open data

---

[5] http://www.cisco.com/c/dam/en/us/products/collateral/wireless/mobility-servicesengine/city_of_barcelona.pdf

[6] https://techcrunch.com/2016/01/20/urban-transport-app-citymapper-snags-40m-from-index-benchmark-yuri-milner-others

initiative generated a value of over £56 million in saved time for users of their services.

Many of the current city data offered in government data catalogues portals are obviously useful. Data produced by services and applications and offered in government data catalogues can be integrated, analysed and visualised. At their best their flows of data can become the differentiation factor for proving new business models and delivering holistic and interoperable digital services. The integration of city systems at the system-of-systems level has been demonstrated to be able to create drivers for infrastructure innovation on and improve the control of resources (Gann et al., 2011).

Despite the considerable potential of city data, the process of designing data catalogues, which offers either static or real time data, has been perceived as complex and cumbersome. City data is provided by a multitude of systems, devices and applications, whose logistical distribution varies according to its suppliers, their sectors, distribution channels, and the policies and regulations to which it is subjected to. Each nature of city data must be tackled in a different way, and the data collections that offer opportunities to make cities smarter must be brought together and become part of a coherent and interoperable whole.

The diversity, instability, and ubiquity of city data make the task of processing, integrating, and interpreting the real world data a challenging task. As it is often difficult to understand the context associated with city data, it is very hard for stakeholders and machines to access and interpret city data unambiguously. Consequently, current smart cities data integration often becomes more about enabling isolated and costly data-driven solutions rather than addressing data integration at the data catalogue level and making the most of standards available for that purpose (e.g. Hypercat[7]).

This confusion may be partially explained by the current global competitiveness on the provision of smart cities solutions demonstrate the extent to which technology providers and disjoint city-led approaches are seeking to shape up the offering of city data. They are in fact, however, exacerbating the problem of data integration for using their set of specific deployments, proprietary technological approaches, data standards and policies.

While some cities have awakened to these risks, they are much less clear on what to do about them and start orchestrating and providing city data through shared and open infrastructures. In fact, the most common city responses to data platforms design have been neither strategic nor outcomes-oriented approach but cosmetic: media campaigns and government publications, which are often composed by numerous non-replicable actions or sub-scale pilots / demonstrators and the showcase of fragmented and "deprived" data re-use.

Existing literature on smart cities and data management rarely offer a coherent framework for data platforms development activities, let alone a strategic one. Instead, they aggregate tales about uncoordinated and disjoint initiatives to demonstrate a city's open data vision. The final outcome is that existing initiatives are susceptible to high development and maintenance costs, information friction, requirements neglecting, non-compliance to data policies and regulations, privacy and data licences violation. These consequences clearly demonstrate the extent to which data platforms are failing the delivery of smart city outcomes, and highlight the potentially large financial and regulatory risks for any city whose data management "conduct" is deemed unacceptable (e.g. misuse of personal data, data license infringement, cyber security / cyber terrorism).

The current proliferation of data catalogues has been paralleled by growth in city data ratings and rakings. While rigorous and reliable ratings might constructively influence cities approach towards the provision of city data, the existing self-appointed scorekeepers does little more than add to the confusion. For instance, the Global Open Data Index measures and benchmarks the openness of data around the world on an annual basis. However, the ranking criteria used in the rankings are not quite convincing. The ranking is based on the analysis of openness of 10 pre-defined datasets only and not on the overall collection of datasets a city catalogue hosts. It weights factors such as the existence of the data, its online availability and formatting. Beyond the choice of criteria and their weightings lies the even more perplexing question of how to judge whether the criteria have been met. The criteria "Is the data machine readable?" is a criteria that is not very easy to measure. Data can be available in digital form but not all digital can be processed or parsed easily by machines.

---
[7] http://www.hypercat.io/

Finally, even if the measures chosen reflect data openness impact, the data are frequently unreliable. The Open Data Institute awarded the London Data Store with the "Open Data Publisher Award", which celebrates high publishing standards and use of challenging data. However, the great majority of data on the London Data Store has no standardised metadata attributes, is out dated, have neither clear licence agreements nor quality assurance. Often, data is machine readable but not understandable. The Open Data Institute (ODI) and University of Southampton's Open Data Monitor project revealed that of the 218 data platforms (European, national, regional, local) they investigated, nearly 50% of the datasets have no standardised metadata attributes, and present 25 different data licence descriptions within their data catalogues. The standards used in data platforms are almost always incompatible with each other, hindering the platform-to-platform integration and the reuse of applications development. In the current moment such indexes tend to use measures for which data are readily and inexpensively available (not pre-processed), even though they may not be good resources for the smart cities outcomes they are intended to support.

In an effort to move beyond this confusion, a growing literature on open data and data platforms has emerged, though what practical guidance it offers to governments is often unclear. Examining the prevailing strategy of data catalogues or platforms design is an essential starting point in understanding why a new approach is needed to integrating both technology and non-technology components more effectively into data management and business models.

## 3. Two Prevailing Approaches in the Provision of City Data

Broadly speaking, the providers of city data in smart cities have used two main strategies to make their case: either to adopt a *bottom-up* or a *top-down* approach. On one hand, bottom-up strategies to city data provision are independent approaches which are often neither addressing the needs of the stakeholders of city data nor integrating their capabilities to city's larger strategies (the top-down approaches). On the other hand, top-down approaches are city-led approaches which are often neither taking social influence into account nor maximizing the efforts of other initiatives who are working towards the provision of city data (the bottom-up approaches).

Bottom-up approaches have been widely adopted in confined and disjoint initiatives. Such initiatives address the problem of city data management as single IT development project which is not fully integrated nor linked together in a way that allows cities to efficiently leverage their collective knowledge. Often bottom-up approaches struggle to maintain interoperable data that is seamlessly integrated across different systems and stakeholders. London's functional bodies for instance encompass the 32 Boroughs, the City of London, the Greater London Authority (GLA), Transport for London (TfL), the Mayor's Office for Policing and Crime (MOPAC), London Fire and Emergency Planning Authority (LFEPA), and the London Legacy Development Corporation (LLDC). All of these groups generate their own data but share little of it with each other. The Infrastructure Mapping Application[8] developed by the Infrastructure & Growth group of the Greater London Authority is one of the emerging public sector solutions challenging this fragmentation. The private sector is an active producer of data in cities. Private sector data are produced, collected or funded by the private organisations (e.g. telecom, utility companies), which can be either released as open or proprietary (commercial) data. In case of proprietary data, it is often represented using industry standards, is subjected to charge, usage authorization, licencing agreements, privacy restriction, and distribution boundaries which are decided by an individual organisation.

---
[8] https://www.london.gov.uk/what-we-do/business-and-economy/better-Infrastructure/londons-infrastructure-plan-2050-progress

As the need for data integration grows, the problem of data interoperability is further exacerbated by the lack of widely-accepted standards for expressing the syntax and semantics of the city data. In such environments incapable of reciprocal operation with other there are high levels of heterogeneity in terms of data, hardware and software elements. Together they become a set of specific technological non-interoperable deployments of components from specific vendors. The implication of this scenario is that each solution/application may interoperable on its very own and original setting. Upgrading this complex environment is time consuming and has high cost implications, particularly if there is a high level use of proprietary and legacy systems (Goldsmith and Crawford, 2014; Suzuki, 2015).

Unfortunately, one of the outcomes of these strategies is the user's frustration at the existence of too many data initiatives and lack of ability to discover the appropriate data. Besides different formats and standards, the data offered in these catalogues present inconstant quality, provenance, and licence agreements. The term data provenance refers to the process of tracing and recording processes that led to the creation of a resource, and can provide additional evidence for accuracy, timeliness and ownership of the data.

Top-down approaches – which suggests that cities act solely as "implementers of initiatives" – are prominent in the goal of the majority of government open data strategies. Led by the smart cities "buzzword", many cities have assumed that answering to governmental pressures for data release before understanding technology and users' needs for data finding, processing and sharing is the best tactic to follow. That may be a natural reaction, especially coming from cities whose data catalogues have been outsourced to private organizations which are willing to sell their proprietary solutions; however, it can be a dangerous one.

By failing to follow appropriate leadership strategies, many initiatives created to offer city data and support smart cities vision haven't been as productive as they could. Most top-down approaches neglect the expectations and needs of the various users of city data (e.g. citizens, businesses, entrepreneurs, data scientists, research institutes, city councils). Data.gov.uk and data.gov are well known for being reasonable proxies for public sector information available to the general public, although certainly not comprehensive sources. As same as the aforementioned bottom-up approaches, the users of these initiatives are not provided with data that can be both human and machine readable and understandable. For instance, nearly 60% of the top ten data formats provided in data.gov.uk are proprietary formats (e.g .pdf and .xls). A very large number of data sets in both initiatives require substantial human workload to data cleansing and semantics sorting (Janssen et al., 2012; Lee, 2012; McLaren and Waters, 2011; O'Riain et al., 2012). Often users are provided with obsolete and non-valid data accompanied by insufficient and poorly documented metadata attributes (Conradie and Choenni, 2012; Jin et al., 2011; Schuurman et al., 2008). User's different competences in data analysis and manipulation is an issue commonly overlooked in both top down and bottom up approaches. It cannot be assumed that the public will have the same amount of knowledge and competences for city data manipulation as researchers and data specialists. Thus, in order to achieve a large-scale dissemination, it makes necessary to lower the knowledge required for data access, and provide means to people to easily discover, reuse, and share data.

The two strategies of city data offering share the same weakness: They focus on the tension between their ambitious targets for data provision and technology rather than on the strategic role of non-technology components in bringing them up together. These approaches even in combination are not enough to tie the strategies of community-led developments (e.g. local authorities, utilities, telecom, and private organisations) to major strategies (e.g. city and national led approaches). Consequently, none of the existing city data strategies on its own is sufficient to help a city to identify, prioritize, and address non-technology and technology issues that matter most or the ones on which it can make the biggest impact in integrating cross-domain value-added city data. The result is oftentimes restricted and uncoordinated city data activities disconnected from the wider smart cities context that neither make any meaningful impact nor strengthen the city's long-term competitiveness in the smart cities arena.

Internally, city data initiatives and practices are often isolated from similar efforts – even separated from their overall vision. Externally, they become diffused among numerous unrelated efforts, each responding to a different stakeholder group or a business model. Taken together, these challenges mean that both isolated top-down and bottom-up data management approaches cannot work. The consequence of adopting such approaches is a tremendous lost opportunity. The power of cities to deliver smart cities outcomes is dissipated, and so is the potential of cities to take actions that would support both their smart cities vision and their strategic and economic goals.

However, overcoming these challenges is not a simple task. Today's simple reality is that it is tremendously difficult for cities to specialise in all the competencies involved in capturing, storing, orchestrating, maintaining and distributing cross-domain city data. Hence, cities can be better served by designing data infrastructures which employs the concept of data driven value networks, leading to the delivery of new services, themselves underpinned by new value chains / networks. Such value networks acknowledge the relationships that make the city data supply chain more effective in a way that produces lower costs, better data and services, and which lower risks for each of its participating members.

## 4. The Link Between Data Platforms and Business Strategies

To advance data infrastructures, we must root it in a broad understanding of the interrelationship between city data offering and business models while at the same time anchoring it in the strategies and activities of specific cities. Although saying broadly that a data platform needs a strategy seems like a very straight forward concept, it is the basic truth and the comprehensive activity that will pull cities out of the confusion that their current data offering approach has created.

Successful smart cities need effective data infrastructures. The simple reality is that cities must respond to the increased market demand for a more integrated provision of city data. Ultimately, a true smart city provides value-add data which expands demand for business and innovation in the long term, as more social needs are met and aspirations grow. Any smart city that pursues its ends based on poor quality data in which its infrastructure will operate on will find its success to be illusory and ultimately temporary.

At the same time, data infrastructures need effective business models. Complex products and dispersed services have been increasingly delivered across many firms and markets, making extremely difficult a firm to innovate alone (Dougherty and Dunne, 2011; Williamson and de Meyer, 2012). This has increasingly led to the disaggregation of firms into complex supply chain networks involving many partners who specialize deeply on their products and services. Regarding smart cities, the provision of city data and specialised services for data manipulation are frequently centrally controlled and excludes units outside the organization boundaries. Nowadays cities, I argue, have neither the tools nor the means of bringing external partners round to the necessary supply chain networks. This is especially due to a number of barriers for joint collaboration beyond organisational barriers, including funding, expertise, and the willingness to work across silos. The consequence of not doing so is that complexity has often overtaken the management of city data. Often, solutions to manage city data have become longer than a simple ICT project, and as a consequence, more expensive and difficult to design and maintain. If smart cities weaken the ability of managed data to address city-wide challenges of joining-up across city silos, they will not allow city data to be exploited to its full effect deliver the smart cities promise.

Leaders of city data offering in smart cities have focused too much on the friction between technology components and not enough on the points of their intersection with non-technology components and business models. The mutual dependence of city data offering and business models implies that any strategic and technology decisions must follow the principle of shared value, or what we call a middle-out leadership pattern. That is, choices must benefit both top-down and bottom-up approaches. If either top-down or bottom-up approaches pursues standards, policies and regulations that benefit only their own interests, it will find itself on a dangerous path. A temporary gain to one will undermine the long-term prosperity of both sides.

Rather than acting solely as "implementers of initiatives", government initiatives must take social influence into account while maximizing the efforts of other stakeholders who are working towards the achievement of the same goal: to facilitate deep knowledge discovery and the creation of new valuable integrated services through the exploitation of rich, interoperable and engaging cross-domain city. Data infrastructures can provide many functions that transcend space (and time), break down the barriers to information access and enhance communication and collaboration. Thus, it enables people to have access to information that will enable them to innovate, to work better, to commute more efficiently in between places, enable governments to get insights on the urban services being provided anywhere and anytime they want.

To put these principles into practice, our business models framework combines city data offering with a business model thinking to renew and extend common innovation and competitive strategies (e.g. Casadesus-Masanell and Ricart, 2010; Chesbrough, 2010; Teece, 2010), and address and address intra- and inter-firm issues such as organisational change, value network design, and innovation management (e.g. Al-Debei and Avison, 2010; Breuer, 2013; Morris et al., 2005; Wirtz, 2011). From a practical perspective, the main purpose of our framework is to allow governments to create, deliver, and capture value through data infrastructures which are designed on the basis of social influence and not authority.

The interdependence between city data and business models takes two forms. First, city data offering needs business models to enable the normal course of its supply chain operation. Second, business models are influenced by external forces, which ultimately impacts the offering of city data. These two forms of interdependencies are explained in the following sections as we introduce our business models framework.

## 5. Designing a business models strategy.

The relationship between city data and business models takes two forms. First, city data offering needs business models to enable the normal course of its supply chain operation. The production of city data can be considered similarly as to logistics in the production process, in which its success is directly related to the effectiveness in managing information flows (Kranton and Minehart, 2001). In city data supply chain, stakeholders combine their data and services into complex supply networks providing integrated products to users and machines which consume and re-use the finished products. This network of partners can potentially bring insights about specialised domains and different application markets that one single city government or organisation developing a data infrastructure for smart cities would struggle to maintain in house. Essentially every activity in a data supply chain affects the delivery of a final product to users, creating either positive or negative impacts. As such, cities must to deeply specialise in their core competence – governance - and decentralise their city data portals and catalogues to create specialised large ecosystem of expert partners. Porter's (1998) value chain analysis provides a means for examining internal processes of supply chains and identifying which activities are best provided by others.

While governments are increasingly aware of the impact of effective city data offering in smart cities, these impacts can be more restrained and variable than many policy makers realize. Data infrastructures depend on the context of the smart city, i.e., they depend for instance on location, culture, available data, smart city vision, local regulations and policies. The same data infrastructure will have very different capabilities or impacts in different locations. A one-size-fits-all approach to data platforms transformation into data infrastructures and simplistic approaches to engage stakeholders of city data with one another are unlikely to work.

Porter and Kramer (2006) demonstrate that opportunities for big impacts of corporate social responsibility (CSR) strategy involves interlinking both *inside-out* – one's firm CSR level perspective) - and *outside-in linkages* - environmental factors affecting the CSR. Borrowing the same rationale to the context of data infrastructures, their strategic design will depend on the interdependence of the inside-out linkages – *the influence that the city data value chain exerts on the smart city vision* – with the outside-in linkages – *the extent to which external forces affects the city data value chain*. Porter (1985) suggests that the activities of a business can be grouped under two headings: *primary activities*, which are those directly involved with the creation and delivery of the product or service; and *support activities*, which feed both into primary activities and into each other. Although he considers support activities as not directly involved in the production process, they have the potential to increase its effectiveness and efficiency.

In the context of data infrastructures, we recommend that cities integrate the data infrastructure supporting activities that are citizen-oriented services, technology, organization, value and governance into our framework, named SMARTify, to orchestrate the primary activities of the city data supply chain and guide their business models. The concepts of the five domains of our business models approach and the data value chain framework part of SMARTify are illustrated in Figure 1.

Using our framework, cities can align the capabilities of their data infrastructures to their smart cities vision and demonstrate how it helps cities to achieve maximum critical success factors, such as the partial lists of examples illustrated here demonstrates. The lack of a time dimension of Porter's value chain model could be particularly significant in the world of fast moving technology development. Our value chain framework addresses this issue by incorporating into the model supporting activities originated from a dynamic business models framework. The value chain illustrated in this figure depicts all the activities necessary for a data infrastructure to manage city data. When cities use the value chain framework they create an inventory of opportunities and operational issues that need to be investigated, assessed, prioritized, and addressed. Logistics in data infrastructures can be defined as the process of strategically managing both forward logistics activities and reverse logistics activities of city data, metadata and flows of information by the city through its data delivery channels to maximize adherence to, and compliance with, current and future smart cities goals.

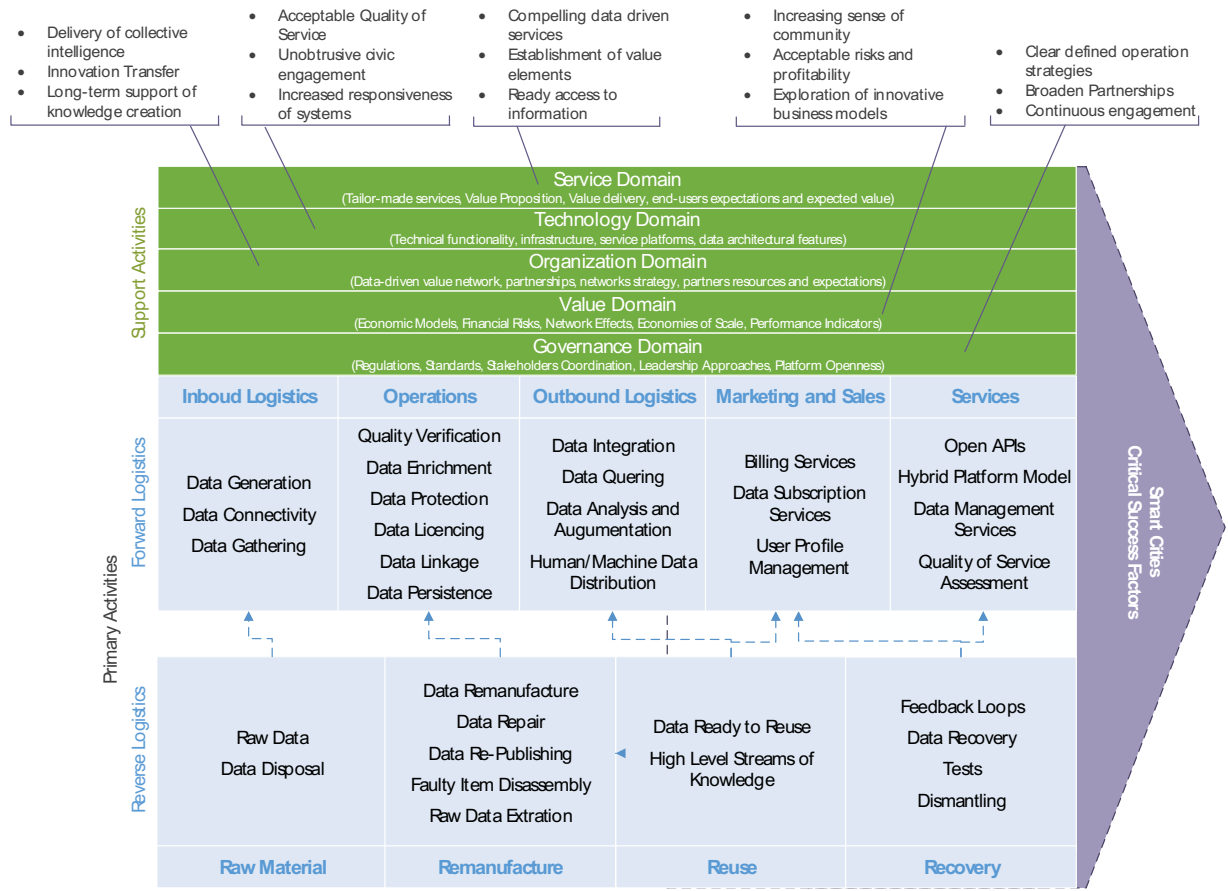

**Figure 1.** SMARTify value chain and business models components: inside-out view.

As an alternative to the prevailing and simplistic use of forward logistics (Chopra and Meindl, 2012) activities adopted in current initiatives, we introduce and align both forward and reverse (Krishnan and Ulrich, 2001) logistics of the city data supply chain. Examples of forward logistics activities to manage city data are acquisition (inbound logistics), handling and storage (operations), distribution (outbound logistics), subscriptions and customer management (marketing and sales), and provision of specialised services (services). In the reverse logistics activities are the collection of feedback (recovery), pre-processing and augmenting (reuse), maintenance (remanufacture), and disposal or raw data extraction (raw material). In Figure 1 we illustrate some basic activities taking place in each stage of the chain, and the set of specialised activities taking place after the data has reached the end of its useful life or is not suitable for usageThe flow of information is vital to the successful of logistics integration in the data supply chain, since all the information (price, cost, market, etc.) must be known to all parties involved in the production process in the same fashion as traditional supply chains. Smart cities' city data supply chain has strong characteristics of a supply network, and is composed by data driven value chain linked by the provision of reusable information and integrated services which reaches with a particular care the last point of the chain that is a satisfied consumer.

Ultimately, our framework helps cities to improve service innovation, technology development, user engagement, and stakeholder's collaboration in data infrastructures. It assists cities mapping out inside-out (smart cities critical success factors) and outside-in (external forces impact ) linkages in their data infrastructure. This framework draws on previous efforts in business models design (Bakry, 2004; Ballon, 2007; Walravens, 1992), value chain analysis (Porter 1985), closed-loop supply chains (Savaskan et al., 2004; Chopra & Mendl 2012), critical success factors (Esteves, 2004), and uses the STOF approach (Bowman, 2008) as the starting point for the creation of a comprehensive business models framework for data infrastructures. To the best of our knowledge, our approach is the unique framework which supports the design and validation of data infrastructure provided by multiple actors. Basically, the SMARTify framework is divided into five broad domains:

1) *Service domain: deliver integrated value propositions*. Support the design of tailor-made data services which careful targets the needs of users and businesses, and explore use cases where data is used to deliver different forms of value.

2) *Technology domain: enable the widespread exploitation of city data*. Support the creation of agreements between stakeholders regarding data handling and technical infrastructure to allow the design of infrastructures that will serve as the foundation for data exploitation in the long-term.

3) *Organization domain: maximize the efforts of stakeholders.* Support the definition of statement of values which city leaders can use to steer a strong value network of collaborators who will provide the expertise needed to deliver a data infrastructure.

4) *Value domain: explore transformative business models and capture impacts.* Support the implementation new and transformational business models that are made possible by increased access to data and closer integration between city systems, and to change existing processes in order to capitalize on these.

5) *Governance domain: deliver impact and accelerate smart cities growth.* Support the clear articulation, measurement, management, delivery and evaluation of the domains and their intended benefits in practice.

The relationships between the concepts within and between the domains are discussed. The five descriptive domain models together provide a descriptive conceptual framework for the design of business models for data infrastructures. The concepts that are most relevant from a design perspective are addressed and illustrated in Figure X as a descriptive conceptual domain model. In the following sections we will discuss these five domains in more detail, and also take a closer look at the theoretical and technological concepts that are the basis for our framework, and that will lead the design of business models.

## 5.1 Service Domain

The main component of the service domain is the costumer value of a product or service. The value proposition of a firm, which is delivered through electronic channels, must be recognized as being better, and as outperforming competitions with regards to human needs and experience (Chen and Dubinsky, 2003; Kotler 1988; Magretta, 2002; Pine and Gilmore, 1999; Huijboom and Broek, 2011). In the case of data infrastructures, cities must engage with citizens and businesses as owners of and participants in the creation and delivery of the data infrastructure, not as outsiders who are merely passive recipients of data and digital services. Ensuring societal needs are recognized as the starting point for city data service offering will be powerful driver of data service transformation.

For this, each smart city must understand the *value expectation* of their users with regards to the data infrastructure, their *previous experience* while interacting with the existing data services provided in the city (e.g. usability, feeling of security, and accessibility) and the *effort* one has to put to utilise these services (Brosch et al., 2012; Economides and Katsamakas, 2006; Ghazawneh and Henfridsson, 2013; Janssen et al., 2012; Williamson and de Meyer, 2012; McDaniel and McLaughlin, 2009). Effort refers to all non-financial effort a user must make to efficiently use data and services offered in the infrastructure. Existing diffusion of innovation literature suggests that service innovations that are perceived by individuals as having greater relative advantage, compatibility, trialability, observability, and less complexity, will be adopted more rapidly (Ilie et al., 2005; Rogers, 1995). Recent research on open data platforms adoption supports these theories (O'Riain et. Al, 2012; Mayer-Schanberger and Zappia, 2011; Le Phuoc, 2009). Reducing human efforts means creating value through lower search and reduce poor data which requires intensive data manipulation, manual processing and verification.

*Table 1. Data Publisher rationale*

| Stakeholder: Data Publisher | |
|---|---|
| Concept | Rationale |
| Context of Use | Publication and management of data |
| Market Segment | Public and private organizations |
| Previous Experience | Difficulty in providing data which complies with established standards, and specifying the conditions and licenses for data reuse. |
| Expected Value | - Easily publish open and private raw data in a machine/ human readable format using agreed standards;<br>- Provide data as a service and charge the release of private data;<br>- Define access-levels to the datasets to avoid data misuse;<br>- Guarantee data is protected with agreed licenses, ownership, and privacy agreements;<br>- Semantically enrich data to make them discoverable by external data repositories; |

A good understanding on *context* of use and the *market segment* of users is also very important to service design. Market segmentation involves dividing the market (of e.g. application developers) into groups with shared needs or desires (Kotler and Wong, 1996), requiring tailor-made data services solutions. A service innovation is only successful when it provides benefits to the costumer in a particular context (Chen and Dubinsky, 2003). A good description of the context in which users would use city data is provided in the work of Davies (2010). Examples of city data context of use are *data-to-fact* (search of specific facts in datasets), *data-to-information (*create static representations or visualisation of one or more data sources*), and data-to-interface (*interactively access and explore one or more dataset). TfL's open data strategy primarily targets the market segment of application developers. Consequently, the context of use TfL mainly supports are *data-to-service*, i.e., transport data feeds are shared via Application Programing Interfaces (API) for a direct reuse of data in mobile, emerging technologies and web applications. A recent rise in user's demand has pressured data.london.gov.uk and data.gov.uk to also provide support to API-enabled datasets. Contexts for city data usage are not limited to these nor are they mutually exclusive and many users of city data employ multiple usage patterns. Table 1 shows an example of a data infrastructure user and its respective rationale.

The value proposition of the data infrastructure puts requirements in the value network or partners who will provide their specialised services in the data infrastructure. Consequently, the service domain is strongly influenced by the *value activities* offered by the value network, as well as the financial arrangements to which their data and services are subjected. Cities will need to explore business cases to define which services are needed on the data infrastructure to enable different modes of data and services usage (see. Korn and Oppenheim, 2011).

Finally, the value activities offered in the platform puts requirements into the technical architecture. They are transformed into technological functionalities and co-determine the value delivered by the data infrastructure. According to Bowman (2008), the *value perceived* by users while interacting with the data infrastructure is defined as the difference between the *delivered value* and *expected value*.

*5.2 Technology Domain*

The Technology domain focuses on the technical architecture which delivers the data infrastructure's technical functionality. Alongside the service design, the technical architecture serves as a guide to the technical design (Bouwman et al., 2008). The technical architecture consists of *applications* to give provide to data and services (e.g. query interfaces, visualisation and data manipulation tools), *service platforms* encompasses for instance data repositories, billing mechanisms, quality of service, service discovery and security mechanisms; and *access networks, devices*, *backbone infrastructure* refers to the medium and long range backbone network infrastructure. On the provision of specialised applications, cities will need to deliver and incentive the development of advance features that will facilitate the collection, management and discovery of data by providers and users. The providers of city data must be provided with data services that will enable them to keep data freshness and follow the general guidance for data publishing.

The technological foundation of the data infrastructure is heavily influenced by the city data and its accompanied meta-data. As the supply and demands for city data growths, the more resources need to be allocated. The starting point to technology design should be performing a city data mapping exercise to produce a picture of the architectural features of city data, including: the sources, volume (Kent, 2000; Ramez and Elmasru, 2010), variety (Clements et al., 2002), temporal factors (Lancy, 2001; Zikopoulos et al., 2012) and sensitivity (Terzis et al., 2005). Table 2 presents a simplistic exemplification of the data landscape in London. This exercise should be followed by the exploration of the vulnerability aspects of city data (e.g. open, private, volunteered citizen's data). Data management capabilities should be developed to ensure data integrity and compliance with National, European and International data protection regulations.Furthermore, data infrastructures ready for the Internet of Things create demand for an ample integration with numerous external resources, such as data storages, services, and algorithms, which can be found within organization units, other organizations, or on the Internet (van Kranenburg, 2011; Perera, et al 2014; Atzori, 2010; Patni et al., 2010). For instance, environmental and infrastructure data (e.g. toxic gases detectors, traffic monitoring, and water quality and leakages detection sensors) are required to be streamed and analysed nearly real-time to trigger systems to act upon emergency situations. Problems in the accessibility and availability of the data will certainly affect the accuracy of the data (e.g. incomplete data transfer), and consequently, decrease the quality of the response. In this context, it is important to identify whether the data from the Internet of Things will be persistently available or how to overcome from a situation where the source of data is unavailable.

An open and hybrid approach model to the delivery of city data seems to be appropriate for the data infrastructure. Whereas the Pipe model provides users with data bulk download, the Hybrid Platform model supports the provision of data *on demand* as it shares features of a platform which provides real time data through API's and bulk download. This model can, therefore, fulfil the requirements of data consumers who want to consume historical data and the ones who want to create real-time applications.

The technology design is a source of *cost* and therefore should support the data infrastructure strategy in the long term. For this, cities will need to rely on a citizen-centric, interoperable, open and innovative vision. This vision can ensure data infrastructures are evolvable and able to accommodate additional functionality at later stage at a fair and transparent cost. For instance, designing data infrastructures as modular-based architecture, which relies on stable and well-defined API's and interfaces, can ensure interoperability between the platform, services and the applications provided by the member of the value network. Ensuring platform openness at interfaces will reduce entry barriers and provide targeting opportunities around the platform due to increased transparency and integration (Gawer and Cusumano, 2002; Schilling, 2010). This will, however, require a significant change on the operational model of the current smart cities data strategies, so that stakeholders can collaborate, disperse data can be reused, and fragmented silos of digital assets and services can be brought together to deliver smart city services. Semantic web technologies such as linked data and ontologies have the potential to interconnect disparate and fragmented silos of information within data infrastructures (Lopez et al., 2012; Le-Phuoc et al., 2012; Heath and Bizer, 2011).

Nevertheless, two classes of interoperability issues often limit the realization of these necessary changes: technical and non-technical barriers to interoperability. On one hand cities must ensure systems and data complies with technical and semantic standards aimed to address interoperability issues. On the other hand, there are several non-technology aspects hindering the effective interoperability of systems and data, including different strategies to manage personal data, to exchange data among different stakeholders, and licenses terminology. In the case of licenses, preference should be given to the adoption of National where possible. In the UK, both national and local data sets are encouraged to license open data under the UK Open Government Licence (OGL v2). Both London.dta.gov.uk and data.gov.uk make use of the UK license framework. Similar strategies have been put in place in Australia (AUSgoal) and France which uses the International (ODbL). Others such as the New York Open Data initiative releases data associated with no common open licence. Yet, these open license frameworks are only applicable in the case of open with no restriction for use and re-use. There are calls for the establishment of licensing frameworks which takes into account various conditions of use and re-use of city data, especially when it comes to proprietary and commercial data. Effective implementation of data infrastructures will need integrated approaches which adopts local, national and international standards where possible. The Standards Hub, IEEE P2413, W3C Semantic Web Standards are example of standards initiatives that can be explored.

*5.3 Organisation Domain*

No data infrastructure provider can solve all of data management problems alone or bear the cost of doing so. Hence, the major component in the organization domain is the establishment of a data-driven value network (DDVN). This particular network consists of several actors and their interactions, each one with its own *strategies*, *goals*, *resources*, *capabilities, level of commitment, influence,* and *requirements*. Their *value activities* are orchestrated in a value chain which enable co-opetition in the provision of city data and services, and ensure response to demand that creates innovation and value. Actors and their value activities alongside with organizational arrangements are combined into roles. Following previous classification of stakeholders (Iansiti and Levien; 2004, Jansen and Finkelstein, 2009; Bowman, 2008) we have defined four basic types of roles in a DDVN: platform keystone, structural partners, contributing partners and supporting partners. They all have varying degrees of power within the value network, based on their resources and capabilities as shown in Table 3.

*Table 2. Sample of London's city data landscape*

| System | Dataset | b/reading | Gb/day | Type | Architecture | Frequency | Sensitivity |
| --- | --- | --- | --- | --- | --- | --- | --- |
| **Metering and Sensing** | Energy Metering | 360 | 49,438 | Sensory | Semi Structured | Near Real-Time | Yes |
| **Metering and Sensing** | Smart Lights | 180 | 52 | Sensory | Semi Structured | Non- Real-Time | Yes |
| **Transport** | Bus Tracking | 200 | 31,311 | Sensory | Semi Structured | Real-Time | Yes |
| **Transport** | Oyster card | 133 | 339 | Text | Structured | Near Real-Time | Yes |
| **Transport** | Traffic Cameras | 25,600 | 160,051 | Image/Text | Multi Structured | Near Real-Time | Yes |
| **Environment** | Smart Bins | 27 | 20 | Sensory | Semi Structured | Real-Time | Yes |

The importance of DDVNs is that it is tremendously difficult for cities to specialise in all the competencies involved in designing, building and maintaining an intelligent data infrastructure. Even powerful organizations like Google and Apple need to collaborate with the various members of their value networks (e.g. developers) in order to provide unique and inventive services and applications to end users. Hence, competitive and co-opetive *interactions* may collaborate for increased participation and commitment within data infrastructures, which are important aspects of collaboration in value networks.

Basically, the success of data infrastructures will be co-determined by the way the value network is managed and nurtured. Cities will need to engage with partners as owners of and participants in the creation and delivery of city data and services, not as passive recipients of services. Instead of simply instigating one-way communication (Davies, 2010; Ferro and Osella, 2012), cities can actively solicit partners to assist in the elicitation of the organisational requirements of data infrastructures.

Collaboration leads to complex interdependencies between partners, because no single partner has formal authority over one another. Every adjustment has to be discussed and jointly agreed (Klein- Woolthuis, 1999). To govern the collaboration, platform keystone, supporting and contributing partners need to agree both formally and informally on how to divide and co-ordinate their activities. On the other hand, there is the risk that strategic interest may induce partners to act against what is agreed upon, and the platform governance design may provide safeguards and legal agreements to create trust between partners to enable open and constructive collaboration among them.

Finally, there are significant gaps in data provision that, in some cases, lead to inertia with certain datasets not be released or conversely, undue attention being given to datasets that are unlikely to generate significant value but have a low cost of dissemination. There are a number of routes to addressing these data gaps. These range from a detailed, regular audit of datasets to improve tracking of current usage. It further reinforces the importance of supporting feedback loops that enable members of the value network to understand the value customers perceive from the data they are offering, and when necessary, to change their strategies so the users of city data and services are satisfied.

### 5.4 Value Domain

A fair division of *costs*, *revenues*, *risks*, *investments* and *value* is required to make the collaboration worthwhile for all the members of a data infrastructure's value network. Data value, also known as data equity, is increasing rapidly as technological innovations take hold. The value of city data can be divided into two categories: Value from *supplying* data and Value from *reusing* data. Value from supplying data can be of two forms. It can be monetary such as the generation of profit through the commercial exploitation of city data, and be in economic terms such increase engagement with constituents and government transparency. Value from consuming data can be obtained for instance from the development of smart cities applications. Within the voluminous amount of city data lies many potentially profitable insights regarding modelling city services performance, spatial aspects of the city, land usage, citizen's mobility and travel behaviours, and trends. The measurement of data value depends on the available organizational arrangements that make a response possible (Basole and Karla, 2011).

*Table 3. Members of the data driven value network*

| Actor | Influence | Main Activities | Examples |
|---|---|---|---|
| **Platform Keystone** | High | Provides the data infrastructure to be used by approved members of the value network, they provide incentives to encourage more participants to join the ecosystem of stakeholders and innovate around the infrastructure. | City government, National Data Providers. |
| **Supporting Partners** | High | Acts as regulators of standards, data provision, personal data management, licenses and commercialisation of data. | Regulatory bodies (e.g. licensing and standardisation), investors. |
| **Contributing Partners** | Medium | Provides services, tools and data. Although less influential than the keystone and supporting partners, they are specialized in different domains, and their presence is essential to deliver and validate the data infrastructure strategy. | Open and proprietary data providers, local councils, academic institutions, business partners, application developers. |
| **Supporting Partners** | Low | Consumes data and services offered by the data infrastructure, creates of business cases, provides of feedback, and creates of knowledge and insights from the city data. | City data end users, data integrator, data scientists, application developers and the media. |

Understanding the different values that can be created through the use of urban data is essential to identify the enablers and the type of data necessary to unlock a specific value. For instance, monetary and good governance values can be unlocked through the release of aggregated data, while innovative services such as apps and new business require a more granular level data that is real/near-real time and with good quality. Competitive advantage is originated from innovative value-added services on top of data, and providing opportunities for innovation and the creation of new businesses through integrated data.

Nevertheless, opening up data is not always free, and there are some potential costs associated with the production and presentation of city data that need to be considered and accounted for. Other *cost sources* include for example technical architecture (e.g. servers, software licenses), value activities (e.g. provision of city data) and general coordination of the value network. There is a substantial commitment and investment on preparing information to be released, purchasing technologies, and upgrading network infrastructure, which all need to be accounted for.

Among the various costs involved in data infrastructures, we highlight the cost of transactions. Transaction costs include the costs of planning, adapting, executing and monitoring task completion. It occurs when a good or service is transferred across a separable interface, for instance when a provider of city data publishes data in the data infrastructure. In some cases, city data is generated in the course of a public and private sector activities, rather than the data being generated or collected as its core activity (Janssen et al., 2012; Gurstein, 2011; Ferro and Osella, M, 2012). This applies, for example, to the crime data contained in the UK Office for National Statistics open data portal, or Transport for London data which arises from its day to day operations. In other cases, data are requested on demand such as the data requested through the "Freedom of Information act". Often it is assumed that such data requires relatively little investment, however, we should also note that cost varies according to how often the data is collected. There are many costs incurred in the stages of the data life-cycle: content curation, processing, storage, aggregation, and anonymization.

To address this source of cost, cities should explore effective ways to recover costs of opening up data for instance by seeking investment sources and creating alliances with the public and private sectors (Miller and Lessard, 2000; Bouwman and Haaker, 2008). Cities could also assist the owners of both public and private data to agree on criteria as which data generated should be made available and explore use cases where data is used to deliver different forms of value. The definition of business models for the open and commercial exploitation of city data (e.g. subscriptions) can generate internal *revenues* and create *revenue* sources. This revenue sources can foster competitiveness and provision of augmented quality data to end users. The commercial exploitation of city data and their funding models are unexplored concepts that at some point cities will need to address. For instance, Copenhagen has partnered with Hitachi to create a data platform for the commercial exploitation of city data provided by the Internet of Things (REF).

The cost structure of data infrastructure may be comparable to cost structure of mobile services, which is characterized by a high-ratio of fixed to variable costs (Shapiro and Varian, 1999) and by a high degree of cost sharing (Guiltinan, 1987). High degree of cost-sharing leads to economies of scope, as the provisioning of a number of different services on a shared infrastructure leads to reductions in cost. Modularity in the service provisioning architecture is a way of obtaining this cost advantage, as components or modules may be shared by several services.

*Risk* sources existing in many domains may have financial consequences. The way the value network copes with the uncertainty and possible financial consequences of the risks needs to be defined. Risks in data infrastructures may arise from performance, privacy and regulatory nonconformity. Furthermore, in the same way actors can become investment sources as they instigate revenue sources (e.g. commercialising data) they can also become risk sources as they can cause strong organisational dependency which may threaten the revenue source.

Equally important to revenue sources are the *network effects* and the *economies of scale*. Network effects or externalities affect the demand-side economies of scale, meaning that the demand of a service or goods defines its value (Shapiro and Varian, 1999). Bowman et al. (2008) suggests that market adoption and usage, both directly influenced by user's perceived value, as well as revenue and return on investment as examples of performance indicators to be considered in the evaluation and management of financial arrangements over time.

### 5.5 Governance Domain

While architecture can reduce structural complexity, governance can reduce behavioural complexity. Platform Keystones must shape and influence its value network, not direct it (Williamson and de Meyer, 2012), besides respecting the autonomy of its members while also being able to integrate their varied contributions into a harmonious whole. This is the essence of platform orchestration which its key function is to provide a context in which distributed innovation driven by value creators can emerge around a platform. It is the platform governance that determines whether innovation divisibility made possible by modular platform architectures is successfully leveraged (Boudreau, 2010; Rochet and Tirole, 2003; Tiwana et al., 2010). Governance of a platform broadly refers to the mechanisms through which a platform owner exerts influence.

The level of openness indicates the degree to which new stakeholders can join the value network and are allowed to drive the strategy or provide services on the infrastructure (for example). The higher the desired level of control and exclusiveness is, the more likely a closed model will be adopted for the data infrastructure. On the other hand, reaching many stakeholders may be an argument in favour of choosing an open model. Cooperation between stakeholders and institutions is very important to develop and to maintain an efficiency data infrastructure for all. Active cooperation can shift the emphasis from the power to decide to the power to transform (i.e., deliver), key to overcoming the delivery deficit of efficient urban services (Morgan, 1997).

Platform keystones make decisions about what expertise should be provided in-house and what is left to supporting and contributing partners. They also retain the power to alter the rights and privileges of users and set contractual obligations and rules of participation. This gives them the flexibility to make changes to the degree of platform openness over time. However, centralizing decision rights with the platform owner may result in risk of overlooking a critical type of complementary knowledge that is likely to be a platform owner's weakness: deep knowledge of user needs. Data infrastructures are multi-sided markets (Gawer, 2008), and therefore, users and members of the value network will have different expectations and requirements with regards the platform.

The value network should be allowed to provide input into platform strategic decisions because they are likely to be able to contribute with distinct types of knowledge that are needed by the platform owner for making decisions. Supporting partners can provide the expertise needed to define policies and actions to address interoperability issues, and decide which standards should be adopted (local, national, international) so that services fit together and that synergies can be exploited. They also identify opportunities that arise to enter complementary markets (e.g. commercial exploitation of city data) and making use of mechanisms they have at their disposal to stimulate innovation within the ecosystem of partners. Contributing partners such as application developers, on the other hand, are closer to and represent the great pulse of emerging end users who will be driving the demand for urban data. Platform keystones may disclose technical and architecture blueprint details in order to share, outsource expertise and partnerships, as well as integrate contributing partners' solutions into the infrastructure itself. We also recommend that cities include public participation and consultation as an active and ongoing process throughout the development of the business models (Gurstein, 2011; Ferro and Osella, 2012).

Establishing and maintaining both competitive and collaborative relationships within the value network may ensure that jointly activities which are based on commonly negotiated terms and conditions can be carried out. The goal behind many partner agreements is the optimization of the data infrastructure operations and services. By entering these agreements, cities can directly benefit from their partner's economies of scale and specialized knowledge, which they could not achieve on their own. Decision makers and providers of data infrastructures should reflect on what kind of partner resources could leverage their business model and their own competencies. For this, optimal platform openness levels and their respective desired control, exclusiveness and target groups of services and data should be defined in the early stage of the data infrastructure development.

Given the strong tradition of internal silo-based approach to city data, cities will need to work across organization boundaries to clearly articulate how participating and collaborating in the value network will benefit all sectors. Furthermore, cities must ensure partners' ongoing participation on the strategy and have a formal managed stakeholder's engagement program in place.

## 6. Data infrastructure's value chain and external forces.

Data infrastructures will operate within a dynamic context, which significantly affects its ability to carry out its business models strategy in the long term. The development process from a business models strategy to established data infrastructure can be divided into a number of phases. In the business models literature, each of this phases are accounted to help understanding the evolution of the competitive landscape and the consequences that such events bring to the firm strategies and business models (Afuah and Tucci, 2001).

In our approach, development process of business models is divided into four phases. It includes the three phases of a dynamic business models defined by Mason and Rohner (2002) and Afuah and Tucci (2001), which was later simplified by Bowman (2008) as Technology/R&D, Roll Out and Market. The three phases involve the conceptualisation of the service, market launch, and the continuous evaluation and improvement to achieve market maturity phases, respectively. We complement the three phases with a forth phase defined as Procurement. Through the business models, cities should be able to identify and procure the best technical solution for their data infrastructure whilst demonstrating the local economic benefit of that procurement (European Commission, 2013). Figure 2 illustrates the four business models phases of the SMARTify framework.

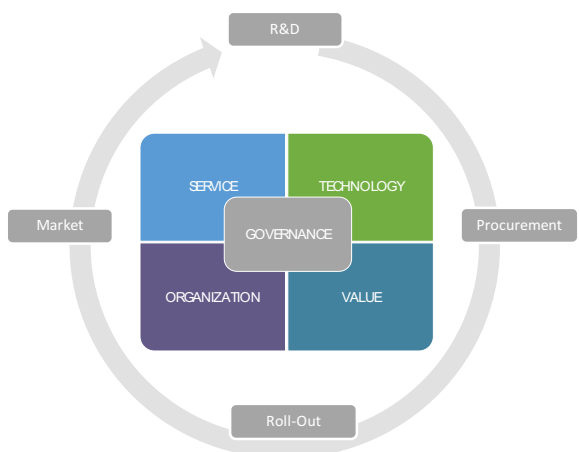

**Figure 2.** SMARTify business models components.

In each of these four phases, business models are influenced by external forces which will ultimately impact the offering of city data. Hence, the business models of data infrastructures are dynamic rather than static, and it is important to understand how they transform over time. These external forces, according to Porter and Kramer (2006), can be referred as to *outside-in linkages*. Hughes (2007) argues that the nature of external constraints on business models can be technical, economic, cognitive, structural, legal, political and cultural. Bouwman and Haaker (2008) defines market drivers, technology, and regulation as external forces with the most direct impact on business models. Our findings suggest that data infrastructures are influenced by feedback loops and the niche players in the value network (Suzuki, 2015).

In addition to understanding the impact of the data value chain on smart cities' critical success factors, effective data infrastructure strategy takes into account the external factors affecting the data infrastructure's competitive context, as illustrated in Figure 3. This figure illustrates the external forces affecting the four business models phases defined in the SMARTify framework. Such external forces affect one's data infrastructure ability to improve operation and execute strategy.

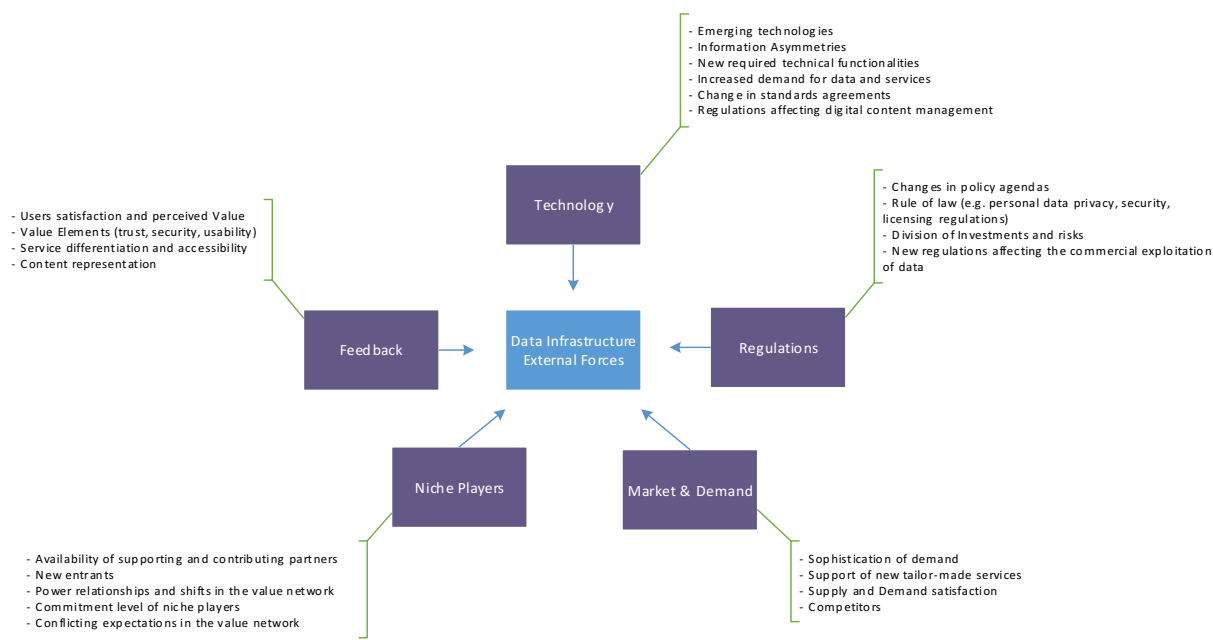

**Figure 3.** SMARTify business models external factors: outside-in linkages.

The impact of these external drivers on internal business model components will be different in each phase. Technology is often the major driver behind new business model and significantly affects R&D phase. For instance, the decision of launching the first version of the London Data Store as a Drupal web portal was based on the easy-to-deploy factor of this technology. This phase can also be affected by market developments, such as observations on the customer demand on existing/new strategies on the data infrastructure market. The procurement phase is mostly affected by changes in technology specifications and regulations. At this stage there will be many design choices and service providers to be aligned with the regulations and specifications of the procurement phase.

During implementation/Roll-out phase there is a risk that strict and changes in regulations can be put in place by regulators or members of the value network. As a result, modifications may be necessary to guarantee that the service complies with such changes. Feedback of users play a major role during the roll-out phase. Early adopters can report on the quality of service and the usability of the services provided. This feedback can be useful to make the necessary changes to meet the customers' expected value.

Furthermore, as technology and services are replaced/extended it may be possible that new partners join the value network introducing new requirements especially in the organization, value and governance domains. Finally, during market phase the focus of data infrastructure providers should be on retaining customers and users, and updating the business models as necessary to ensure their strategies are still competitive and the business offering is still positively supporting the delivery of smart cities outcomes.

## 7. Assessing the Viability of Data Infrastructures.

At the heart of any data infrastructure strategy is a unique value proposition: support the smart cities outcomes that the city can meet for its citizens that others cannot. The most strategic data infrastructure design occurs when cities aligns its value proposition to the outcomes of the smart city while taking into account the design variables that are crucial to the viability of the overall strategy. While the SMARTify model in itself serves well as a holistic business model design of the data infrastructure, the viability of this model can be assessed based on understanding the relationship between the data infrastructure's critical design issues (**CDI**) and critical success factors (**CSF**).

- **CDI** are potential common and recurrent critical design variables which are crucial to the viability and sustainability of the business model. CDIs are defined for each domain of the SMARTify approach, and they vary accordingly to the context and services provided by a data infrastructure.

- **CSF** Critical factors are those, which are essential for successful implementation of a business model. The identification of such factors may encourage their consideration when cities are developing an appropriate implementation plan as seen in industry (Mann and Kehoe, 1995).

In contrast to existing frameworks (e.g. Giffinger, 2007), which attempts to define a standardised framework to rank smart cities which is not associated to individual city needs and strategies, our framework we suggest cities to define CSFs as the ones that will support the realisation of the capabilities of smart cities shown in Table 4. This table summarize the definitions of each urban capability which will be used to formulate design propositions that describe the relationships between the CDI and CSF.

We argue that each city has its own political, cultural, technological and governance settings, and therefore, they must specify the strategy that suits their needs and context. For instance, the deployment of electronic panels at bus stops may not be necessarily considered as an indicator of "smarter transport" in every city. Most ratings rely on different indicators and public attention resulted from marketing campaigns which provides subjective and deviant final ratings, as also observed in (Schönert, 2003, Hollands, 2008).

Hence, in our approach we've found sensible to provide cities with a tool that will enable them to define, assess, validate and evolve their urban capabilities in the long-term. The Oxford dictionary defines capability as *"a valuable resource of a particular kind"* while The Free Dictionary defines it as "assets available for use in the production of further assets". Through the evolution of urban intelligence, we have notices that the development of a particular capability enhanced and produced new capabilities which are the essential foundations of the cities of the future.

The assessment of the business models viability is performed in three interdependent steps, as illustrated in Figure 4. The first step consists of identifying the critical design issues of the data infrastructure. Critical design issues for data infrastructures can be defined based on industry and academic materials and case studies, as well as previous data strategies and public consultation. Recurrent issues and their perceived relevance with regards to the viability of the business model can be qualified as critical.

Once specific CDI are identified for every domain, the second step of the business models assessment can be initiated. This step consists in systematically clustering the CDI and using them to assess and balance the requirements specifications elicited in the business models analysis. In the case any requirement specification negatively impacts a critical design issue, a requirements trade-off analysis must be carried out. For instance, consider the simplistic example shown in Table 5, in which a data infrastructure must support the federation of data from external data sources, and at the same time satisfy pre-defined critical design issues, such as Target Users (service), User engagement (service), Interoperability (technology), Broaden Partnership (organisation). On one hand, increased content targets and engage users with the data platform as well as increase the partnership with external data providers (contributing partners). On the other hand, federating data from other datasets significantly compromise data interoperability and requires the implementation of several mechanisms to mitigate semantic mismatch. Based on such arguments, this requirement should not be satisfied at the moment and revisited at a later stage when circumstances change.

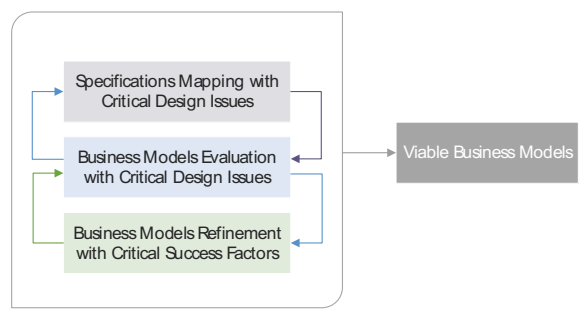

**Figure 4.** SMARTify assessment of a viable business model.

The third step consists in validating the CSF based on the CDI which are instrumental in the delivery of the data infrastructure's value proposition. They serve as the starting point for the design of the CSFs causal models. We have created a break-down structure that explains how business model viability can be influenced by more concrete,
design-oriented variables that influence the success factors regarding creating urban capabilities. For instance, a number of CDI influences the CSF associated with creating human capability in smart cities. Figure 4.11 illustrate the CSFs for Creating Human Capability. In this figure the grey boxes represent the CSF, while the boxes shown in colour refers to the CDI identified for each one of the five domains of the proposed business models (see the legend at the top right corner of the figure).

The CSF *Clearly Defined Target Group* is enabled by the provision of services that support the multi-context data usage of users (e.g. data-to-information, data-to-services). The CDI Value Elements (e.g. trust, security), Reputation and Pricing (finance domain) enables the creation of a second CSF *Compelling Citizen's driven services*. Reputation can be used to differentiate the value proportion from those of competitors (Kotler, 2000), and is influenced by the partners of the value network. Through the adoption of efficient mechanisms to follow up on civic engagement (e.g. assessment of value perceived and collection of feedback) the data infrastructure can be continuously improved to respond to user's demands. As a consequence, the CSF *Increasing participation* in content use and reuse, and take up from users can be achieved (Hey and Trefethen, 2003; Janssen et al., 2012). The CSF *Acceptable Quality of Service* is influenced by the quality of the services delivered, the security in services, data communication and distribution, and the discoverability of contents and easy-of-use aspect of the data infrastructure CDI. Technology aspects strongly influence the services delivered and together they should lead to an acceptable quality level.

*Table 5. Mapping Requirements Specification to CDI*

| Requirement 1 | The data infrastructure shall support the federation of data from external data sets | |
|---|---|---|
| **Priority** | Must have | |
| **CDI** | **Effect** | **Rationale** |
| **Target Users** | + | Users are offered data that will suit their needs |
| **User engagement** | + | Users engagement is increased as content grows |
| **Interoperability** | - | Data interoperability is compromised due to significant semantic mismatch |
| **Broaden Partnership** | + | Content is increased through the creation of partnerships with external providers |
| **Solution** | A great range of interoperable data is desirable but federating data may compromise the interoperability of the data and should be avoided at this time. Change priority to **could have**. | |
| **Criteria** | Satisfying this requirement will require significant extension of data processing capabilities in the data infrastructure. | |

Table 6. CSF Assessment of the Data Infrastructure

| Capability | CSF | Status | Assessment |
|---|---|---|---|
| **Human** | Clearly Defined Target Group | Positive | The service focuses on providing high quality open data and specialised services to the stakeholders of smart cities |
| | Compelling citizen's driven services | Positive | Through the provision of trustable, easy-of-use and high quality data and services the platform is able to provide services that will satisfy the demand. |
| | Increasing Participation | Positive | The provision accessible interfaces to collect feedback from users will enable the continuous improvement of services. |
| | Acceptable Quality of Service | Negative | Scalability issues poses a threat to the validity of our business models. The specified infrastructure will not scale if either supply or demand increases by 25%. Scalability should be regarded as an important CDI during the business models analysis when requirements are elicited. |

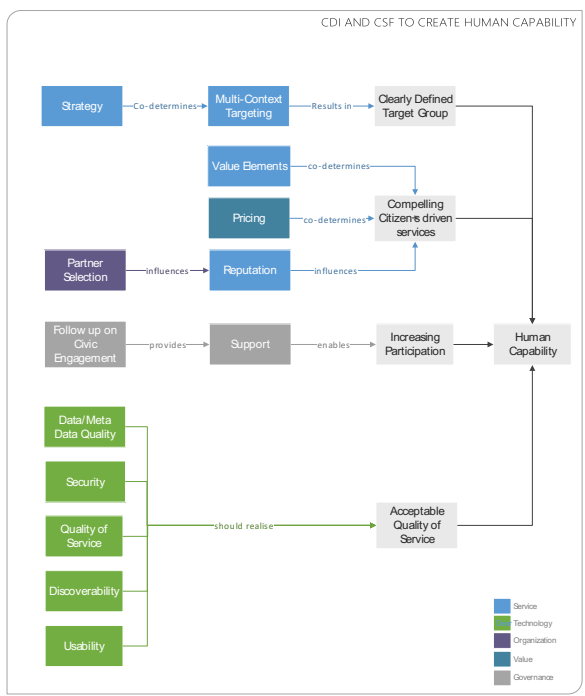

**Figure 5.** Relationship of CDI and CDF in creating human capability.

Where a particular design issue obstructs the accomplishment of a critical success factor it is balanced iteratively until a viable business model is created. Table 6 presents an example of the final assessment of the CSF. In this example, the status – in our case the provision of support to the delivery of smart cities outcomes - of the Acceptable Quality of Service CSF is negative. In this example the critical design issue *Scalability* was overlooked in the design phase. For instance, the data model specification may become a bottleneck in the performance of the data infrastructure as supply and demand increases. In this case, scalability issues should be considered during the business models analysis which is when requirements specifications are defined. Such issues may also arise at times when external factors affect the business models strategy. Our approach considers that positive scores on the CSF will result in a service that supports the human capability of smart cities.

Strategy is always about making choices, and success in managing data as any other city infrastructure is no different. Our findings suggest, that designing data infrastructures based on shared value – the proposed middle-out leadership pattern - should be regarded as a long-term investment in a city's vision to meet future challenges. Cities that make the right design choices and build focused, proactive, and integrated data infrastructures initiatives aligned with their core strategies will increasingly enhance and transform government services and stimulate innovation in city services to the benefit of everyone.

## 8. Related Work

Thus far the quest toward new government business models for smart cities remains problematic. There is a need to develop and apply business models to continue the progress towards creating smart cities and accomplish citizens-orientation. Yet the exact link between e-business models and smart cities initiatives is yet to be explored.

In the current electronic business models literature there is no clear consensus of what constitutes business model for government digital strategies, and thus no established general business models framework. Based on a study of 59 e-government websites, Janssen et al. (2008) classified eight e-government business models, which are based on the eight atomic business models (Weill and Vitale, 2001), but adapted for e-government. These business models are (a) Content provider, (b) Direct-to-customer, (c) Value-net-integrators, (d) Full-service provider, (e) Infrastructure service provider, (f) Market, (g) Collaboration, and (h) Virtual communities. Janssen et al. (2008) found a number of elements that are present in such e-government business models, such as elements to fulfil the mission successfully using the Internet, and to satisfy citizens and/or businesses; organizational network and relationships with other agencies that target the same audiences.

While high-quality experiences with responsive, integrated, Web-based services in the private sector have led citizens to expect the same from the public bodies and agencies (Hazlett and Hill, 2003), how governments can harness Web-based business models to improve their digital strategies that will enable the creation of smart cities remains relatively unexplored in the literature. However, some authors have begun to deal with business models applied to related strategies, such as e-Government business models components (Janssen et al., 2008), e-government initiatives (Bakry, 2004), and mobile services (Bouwman and Haaker, 2008).

In addition to classification of business models, some frameworks for more detailed analysis of e-government business models can be found in academic literature. Bakry has defined a STOPE model for e-government initiatives. It consists of five domains for e-government application business modeling, namely Strategy, Technology, Organizations, People and Environment (Bakry, 2004). Partially based on the STOPE model and building further, Esteves and Joseph's construct EAM (E-government assessment framework) a three-dimensional framework for the assessment of e-government initiatives, based on maturity level, stakeholders, and STOPE domains (Esteves and Joseph, 2008). Ballon (2007) proposes a holistic business modelling framework called Business Model Matrix that is centred around value network, functional architecture, financial model, and the value proposition parameters that describe the product or service that is being offered to end users. However, its business models are relevant to closed systems approach in which the public component is outside the value network. Walravens (Walravens, 1992) extended the Business Model Matrix to support mobile services in cities and have governance and public value as two fundamental elements. However, all these models remain on a high level and provide little or no help in the actual service design process of government digital strategies, and therefore their application to smart cities is limited.

Based on business models and business model frameworks previously developed, Bouwman and Haaker (2008) introduce a holistic model for describing the business models of electronic services, called the STOF model. Similarly to the STOPE method, STOF hides complexity of many other models into four core components, or domains: namely Services, Technology, Organization and Finance. The STOF model focuses on customer value of a mobile services and supports business model dynamics as well, because the model can be iterated in different product phases. Unlike some other business model frameworks such as Osterwalder's business model canvas (Osterwalder and Tucci, 2005), the STOF method takes into account techno-economic interdependencies.

While previous business models frameworks treat business models mainly as a mediator between technologies, strategies, and economic value (e.g. (Chesbrough and Rosenbloom, 2002; Chesbrough, 2010; Hamel, 2000; Johnson and Kagermann, 2008; Teece, 2010), the question of how business models can support smart cities and their stakeholders, and their innovations in creating, delivering, and capturing economic, social, and sustainability value has so far received little attention. To fill out this gap, our approach outlines a new analytical framework whose major purpose is to facilitate strategic-decision making by governments when linking their data platforms with business strategies in order to create data infrastructures.

25## 9. Conclusions

This paper concentrates on the definition of data infrastructures and their business models and components. Our method employs a business models-driven approach to support the elicitation and modelling of requirements and data strategies, and a closed loop supply chain model to serve as a reference architecture model for data infrastructures. By using critical design issues and critical design factors, the positive and negative contributions that may occur among the requirements and specific design needs can be easily identified, as well as the final contributions of the data infrastructure to the realisation of smart cities. Our framework facilitates the requirements elicitation process from business models analysis, as well as the detection of requirements mismatch across the five domains of the business models. It offers templates for requirements balancing and refinement which can be used to determine the trade-offs to be made during the design of such large interconnected systems.

The dynamic business models approach enables decision makers to evaluate the evolution of the business models and how external factors may impact the several stages of the development process of digital strategies of cities. The closed loop supply chain models can give government and their contributing and supporting partners the ability to better collaborate on the basis of common, accurate, reference architecture to design and build more sustainable and cost effective data management solutions for smart cities.

Data infrastructures have a profound and positive influence on the smart cities vision set by cities around the world. The moral purpose of data infrastructure is to contribute to a create a prosperous and sustainable economy. Governments and technology corporations often forget this basic truth. When governments disregard the efforts made by local initiatives within their cities, they penalize productive clusters of stakeholders of smart cities who are working towards the same goal. Such governments and clusters of bottom-up approaches are fated to financial loss, information asymmetry, standards and regulations nonconformity, and hindering the city's ability to leverage its collective knowledge.

Our framework provides to governments and supporting and contributing partners with the clarity they need to change this scenario. It, however, does not intend to describe a one-size-fits-all model for data infrastructures in smart cities. Rather, the focus is on the enabling processes by which innovative use of technology and data coupled with governance strategies, a strong value network of partners can help deliver the various visions of data strategies for cities in more efficient, aligned and effective ways.

Efforts to find shared value in data infrastructures have the potential not only to foster economic and support the delivery of smart cities outcomes development but to change the way government and independent clusters of initiatives relates to each other.

We understand that it will require dramatically different thinking in governments. We are convinced, however, that data infrastructure will become increasingly important to the success of the smart cities strategies. Governments cannot specialise in all the capabilities required to deliver data infrastructures and solve all city data integration problems, nor do they have the resources to solve them all. Each member of the data infrastructure value network can identify the particular set of problems that it is best equipped to help resolve and from which it can gain the greatest competitive and economic benefit.

Supporting smart cities initiatives by creating a shared data infrastructure will lead to self-sustaining and innovative solutions that will create the cities of the future. When cities orchestrate in a value network the stakeholders of their many dispersed city data initiatives, they apply its vast resources, expertise to problems that it understands and in which it has a stake, and cities can take advantage of unprecedented insights into how the city and its infrastructure functions and be ready to overcome social challenges.